\documentclass[journal]{IEEEtran}

\usepackage{graphicx}
\usepackage{epsfig}
\usepackage{dcolumn}
\usepackage{bm}
\usepackage{amsmath}
\usepackage{amssymb}
\usepackage{fancyhdr}

\begin{document}
%
\title{Practical approach to programmable analog circuits with memristors}
%
%
%

\author{Yuriy~V.~Pershin and Massimiliano~Di~Ventra
\thanks{Yu. V. Pershin is with the Department of Physics
and Astronomy and USC Nanocenter, University of South Carolina,
Columbia, SC, 29208 \newline e-mail: pershin@physics.sc.edu.}
\thanks{M. Di Ventra is with the Department
of Physics, University of California, San Diego, La Jolla,
California 92093-0319 \newline e-mail: diventra@physics.ucsd.edu.}

\thanks{Manuscript received August XX, 2009; revised November YY, 2009.}}

%
%


\maketitle

\begin{abstract}
We suggest an approach to use memristors (resistors with memory)
in programmable analog circuits. Our idea consists in a circuit
design in which low voltages are applied to memristors during
their operation as analog circuit elements and high voltages are
used to program the memristor's states. This way, as it was
demonstrated in recent experiments, the state of memristors does
not essentially change during analog mode operation. As an example
of our approach, we have built several programmable analog
circuits demonstrating memristor-based programming of threshold,
gain and frequency. In these circuits the role of memristor is
played by a memristor emulator developed by us.
\end{abstract}


%
\IEEEpeerreviewmaketitle

\section{Introduction}

\IEEEPARstart{T}{he} recent experimental demonstration of
resistivity switching in TiO$_2$ thin films \cite{Strukov2008-1}
and the establishment of a link between this result and more than
thirty-year-old theoretical description of memristors
\cite{Chua1971-1} (resistors with memory) have attracted a lot of
attention to this exciting field. Memristive behavior is found in
many different systems
\cite{Strukov2008-1,Chua1971-1,Chua1976-1,Pershin2008-1,Driscoll2009-1,Driscoll2009-2,Pershin2009-1,Wang2009-1,Liu2000-1,
Yang2008-1,Jo2009-1,Stewart2004-1,Erokhin2007-1,Choi2009-1,Gergel2009-1,Chen2009-1,Oka2009-1,Chen2009-2,Standley2008-1,Rose2007-1}.
To interpret the experimental observations and predict a circuit
behaviour, a number of theoretical models were developed
\cite{Chua1971-1,Chua1976-1,Pershin2009-2,DiVentra2009-2,Strukov2008-1,Strukov2009-1,Strukov2009-2,Cagli2009-1,Joglekar2009-1,Benderli2009-1,Biolek2009-1,Biolek2009-2}
including SPICE models
\cite{Benderli2009-1,Biolek2009-1,Biolek2009-2}. Memristors offer
a nonvolatile memory storage within a simple device structure
attractive for potential applications in electronics including the field of neuromorphic circuits as well, namely circuits which
mimic the function and operation of neural cell networks in
biological systems~\cite{Pershin2009-3,Snider2008-1}. Until now,
most of the potential applications that have been proposed for
these systems have relied on a binary mode of operation (on and
off states of a memristor) while the understanding that memristors
can be used as truly analog memory elements is only emerging
\cite{Pershin2009-3,Pershin2009-2,Delgado2009-1}.

A weaker interest in analog applications of memristors can be
partially justified by the perception that TiO$_2$ thin films
behave as ideal memristors. According to Chua's definition
\cite{Chua1971-1}, the internal state of an ideal memristor
depends on the integral of the voltage or current over time.
Therefore, the use of ideal memristors as analog elements in,
e.g., programmable analog circuits seems to be limited since their
internal state, once programmed, would change significantly due to
a dc component in current or applied voltage. Only {\em perfect}
ac signals would not significantly change the memristor state so
that the use of such devices seems to appear narrow. However,
experimentally realizable memristors \cite{Yang2008-1} are {\em
not} ideal.  In fact, these devices belong to the much more
general class of memristive systems \cite{Chua1976-1} (see the
definition below) allowing for a more complex behavior, which is
at the basis of our proposal.

At this point, a note should be made about the commonly used terminology and the
terminology used in this paper. It appears that in the recent
literature the term memristor has been used for both {\em ideal} memristors
\cite{Chua1971-1} and memristive devices and systems
\cite{Chua1976-1}. Indeed, we expect all experimental realizations of
such devices to be not ideal. Therefore, there is no reason for two
different names. This convention is used in the present paper, so that the
term memristor will refer to {\em all} memristive systems and ``ideal memristor'' will be
understood only in the sense of the definition in Ref. \cite{Chua1971-1}.

In the present paper, we suggest an approach to use resistors with
memory in analog circuits based on threshold-type behaviour of
experimentally studied solid state memristors
\cite{Liu2000-1,Yang2008-1}. Our main idea is to use low voltages
in the analog mode of operation and high voltage pulses in order
to program the memristor's state. In this way, we obtain a circuit
element whose mode of operation is close to that of a digital
potentiometer but its realization is much more simple. Our scheme
represents an important application of a new class of emerging
systems collectively called memory-circuit
elements ({\em memelements})~\cite{DiVentra2009-2}. Therefore, it may be of interest
to scientists from such diverse disciplines as electrical
engineering (in particular, from the area of tunable resistance
research \cite{Ozalevli2008-1,Wee2008-1}), physics, materials
science, and even neuroscience.

This paper is organized as follows. Our approach to use memristors
in programmable analog circuits is introduced in Sec. \ref{sec1}. In
Sec. \ref{sec2}, we present a memristor emulator - an electronic
circuit whose response is similar to that of a memristor. We will
use this circuit in Sec. \ref{sec3} to demonstrate several
applications of memristors in analog circuits including programmable
threshold comparator, programmable gain amplifier, programmable
switching thresholds Schmitt trigger, and programmable frequency
relaxation oscillator. Concluding remarks are given in Sec.
\ref{sec4}.

\section{Definitions and main concept} \label{sec1}

\subsection{Circuit elements with memory}

Before describing our approach, let us give formal definitions of
memristors and briefly discuss their properties. To this end, we
start from the general definition of circuit elements with memory
which include also memcapacitors and meminductors
\cite{DiVentra2009-2}. Let us then introduce a set of $n$ state
variables $x$ that describe the internal state of the system. Let
us call $u(t)$ and $y(t)$ any two input and output variables
\cite{DiVentra2009-2,Chua2003-1} that denote input and output of
the system, such as the current, charge, voltage, or flux. With
$g$ we indicate a generalized response function. We then define a
general class of $n$th-order $u$-controlled memory devices as
those described by the relations~\cite{DiVentra2009-2}
\begin{eqnarray}
y(t)&=&g\left(x,u,t \right)u(t) \label{Geq1}\\ \dot{x}&=&f\left(
x,u,t\right) \label{Geq2}
\end{eqnarray}
where $f$ is a continuous $n$-dimensional vector function, and we
assume on physical grounds that, given an initial state $u(t=t_0)$
at time $t_0$, Eq.~(\ref{Geq2}) admits a unique
solution.

Memcapacitive and meminductive systems are special cases of
Eqs.~(\ref{Geq1}) and~(\ref{Geq2}), where the two constitutive
variables that define them are charge and voltage for the
memcapacitance, and current and flux for the meminductance. The
properties of these systems can be found in
Ref.~\cite{DiVentra2009-2}. In this paper we will instead be
concerned with a third class of memory devices - memristive
systems. Using Eqs.~(\ref{Geq1}) and~(\ref{Geq2}), we can define
an $n$th-order voltage-controlled memristive system as that
satisfying
\begin{eqnarray}
I(t)&=&R^{-1}_M\left(x,V_M,t \right)V_M(t) \label{Condeq1}\\
\dot{x}&=&f\left( x,V_M,t\right) \label{Condeq2}
\end{eqnarray}
where  $x$ is a vector representing $n$ internal state variables,
$V_M(t)$ and $I(t)$ denote the voltage and current across the
device. The quantity $R_M$ is a scalar, called the memristance (for
memory resistance) and its inverse $R^{-1}_M$ is called memductance (for memory conductance).

Similarly, an $n$th-order current-controlled memristive system is
described by
\begin{eqnarray}
V_M(t)&=&R\left(x,I,t \right)I(t) \label{eq1}\\
\dot{x}&=&f\left(x,I,t\right) \label{eq2}
\end{eqnarray}
A charge-controlled memristor is a particular case of Eqs.
(\ref{eq1}) and (\ref{eq2}), when $R$ depends only on the charge,
namely
\begin{equation}
V_M(t)=R\left( q\left(t \right)\right)I(t), \label{eq3}
\end{equation}
with the charge related to the current via time derivative:
$I=dq/dt$.

Several noteworthy properties can be identified for
memristors~\cite{Chua1976-1}. For instance, these devices are
passive provided $R^{-1}_M\left( x, V_M ,t\right)> 0$ in Eq.
(\ref{Condeq1}), and do not store energy. Driven by a periodic
current input, they also exhibit a ``pinched hysteretic loop'' in
their current-voltage characteristics. Moreover, a memristor
behaves as a linear resistor in the limit of infinite frequency
and as a non-linear resistor in the limit of zero frequency,
assuming Eq.~(\ref{Condeq2}) admits a steady-state solution. The
reason for this behavior is due to the system's ability to adjust
to a slow change in bias (for low frequencies) and the reverse:
its inability to respond to extremely high-frequency oscillations.
We will explicitly demonstrate these properties with our memristor
emulator presented in Sec.~\ref{sec2}.

\subsection{Memristors in programmable analog circuits}

Our main idea of using memristors in analog circuits is based on
the following observation of the experimental results: the rate of
memristance change depends essentially on the magnitude of applied
voltage \cite{Liu2000-1,Yang2008-1}. At voltages below a certain
threshold, the change of memristance is extremely slow, whereas at
voltages above the threshold, $V_T$, it is fast. Therefore, we
suggest to use memristors in analog circuits in such a way that in
the analog mode of operation (when the memristor performs a useful
function as an analog circuit element) only voltages of small
magnitude (below the threshold) are applied to the device, while
higher-amplitude voltages (above the threshold) are used only for
programming. The programming voltages can be applied in the form
of pulses. Each pulse changes the resistance of memristor by a
discrete amount.

\begin{figure}[bt]
 \begin{center}
 \includegraphics[angle=0,width=4cm]{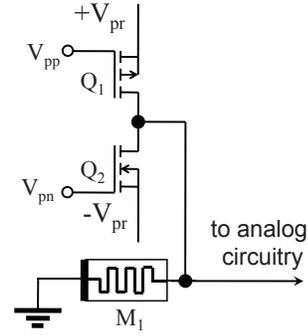}
\caption{\label{fig1} Memristor-based digital potentiometer
consisting of the memristor M$_1$ and a couple of FETs $Q_1$ and
$Q_2$. The external control signals $V_{pp}$ and $V_{pn}$ are used
to set (program) the resistance of memristor $R_M$ between two
limiting values $R_1$ and $R_2$. Here, $V_{pr}$ is the memristor's
programming voltage that should exceed the threshold voltage of
memristor.}
 \end{center}
\end{figure}

In this way, programmable memristors operate basically as digital
potentiometers. However, there are several potential advantages of memristor-based
digital potentiometers over the traditional ones. In particular,
the size of a memristor can be very small, down to
30$\times$30nm$^2$ \cite{Snider2008-1}, allowing for higher
density chips/smaller electronic components. The operation of
memristor-based digital potentiometers requires less transistors
since the information about resistance is written directly into a
memristive medium. Finally, the resistance is remembered in the
analog form potentially allowing for higher resolution. Recent
examples of integration of TiO$_2$ memristors with conventional
silicon electronics \cite{Borghetti2009-1} and of a silicon-based
memristive systems \cite{Jo2009-1} have demonstrated the practical
feasibility of integrated memristor-based electronic components.

Fig. \ref{fig1} shows a simple memristor-based digital potentiometer
which can operate as part of an analog circuit as we demonstrate
below. For the sake of simplicity, one of the memristor's terminals
is connected to the ground while another terminal is connected to
analog circuitry and a pair of field effect transistors (FETs) that
are used to program the memristor state. We use two external control
signals $V_{pp}$ and $V_{pn}$ to open/close the FETs when needed.
When one of these FETs is open, the programming voltage $\pm V_{pr}$
exceeding the threshold voltage of memristor $V_{T}$ is applied and
memristor's resistance changes in the direction determined by the
applied voltage sign.

For definiteness, let us assume that the application of positive
voltage increases $R_M$ and application of negative voltage
decreases $R_M$. Then, the following protocol of programming can
be employed. Since, at $t=0$ we start with a possibly unknown
value of $R_M$, the latter can be driven into the $R_{max}$ state (state of maximum
resistance)
by connection of memristor M$_1$ to $+V_{pr}$ (using the ``on''
state of Q$_1$) for sufficiently long time (this time should be
selected longer, or at least equal to the time required to switch
$R_M$ from $R_{min}$ - state of minimum
resistance - to $R_{max}$). After that, a connection of
M$_1$ to $-V_{pr}$ (provided by the ``on'' state of Q$_2$) for a
specific amount of time will switch $R_M$ into a desired state.
These manipulations can also be performed by application of a
certain number of positive and negative fixed-width pulses. In
fact, we have recently used pulse control of memristors in
memristive neural networks \cite{Pershin2009-3}.

The physical properties of TiO$_2$ memristors were reported and
discussed in several
papers~\cite{Strukov2008-1,Yang2008-1,Strukov2009-1}. The first
theoretical model of TiO$_2$ memristors was suggested already in
Ref.~\cite{Strukov2008-1}. This model does not include any kind of
threshold-type behaviour and simply assumes that the resistance is
proportional to the charge flown through device. However,
subsequent work~\cite{Yang2008-1} has clearly shown that the
charge-based model fails to describe the experimental results. In
particular, Fig. 3b of Ref. \cite{Yang2008-1} shows that a
switching of memristor state occurs at high voltages while at low
voltages sweep-down and sweep-up curves coincide. A first
activation-type model explaining this feature was proposed by us
\cite{Pershin2009-2} and has appeared in October 2008 in the
cond-mat arxive \cite{Pershin2009-2a}. A later publication
\cite{Strukov2009-1} (in early 2009) explains the activation-type
behaviour of TiO$_2$ memristors by a non-linear dopant drift in
which, at low voltages, the change in memristor state is
exponentially suppressed. These types of memristor behaviour are
precisely those needed for our proposal of analog programming.

\section{Memristor emulator} \label{sec2}

\begin{figure}[tb]
 \begin{center}
 \includegraphics[angle=0,width=6.5cm]{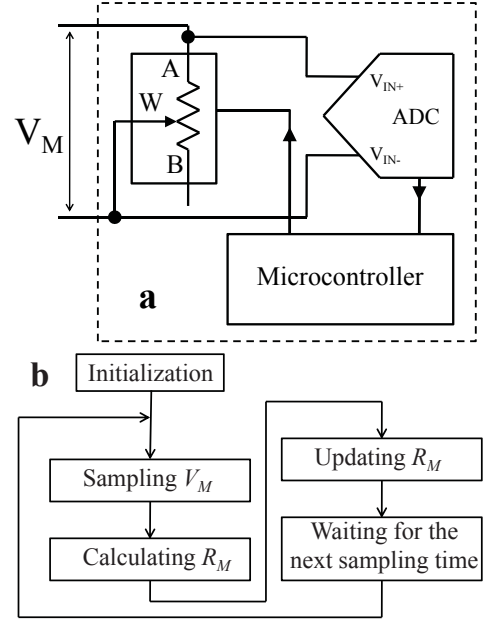}

\caption{\label{fig2} (Color online) {\bf a} Memristor emulator
consisting of the following units: a digital potentiometer, an
analog-to-digital converter and a microcontroller. The A (or B)
terminal and the Wiper of the digital potentiometer serve as the
external connections of the memristor emulator. The resistance of
the digital potentiometer is determined by a code written into it
by the microcontroller. The code is calculated by the
microcontroller according to Eqs. (\ref{Condeq1}) and
(\ref{Condeq2}). The analog-to-digital converter provides the
value of voltage applied to the memristor emulator needed for the
digital potentiometer code calculation. The applied voltage can be
later converted to the current since the microcontroller knows the
value of the digital potentiometer resistance and a
current-controlled memristive system can be realized. In our
implementation, we used a 256 positions 10k$\Omega$ digital
potentiometer AD5206 from Analog Device and microcontroller
dsPIC30F2011 from Microchip with internal 12bits ADC. {\bf b}
Block scheme of the memristor emulator's algorithm.}
 \end{center}
\end{figure}

\begin{figure}[tb]
 \begin{center}
 \includegraphics[angle=0,width=6.5cm]{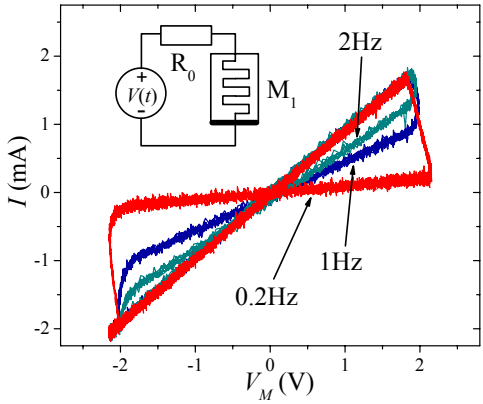}
\caption{\label{fig2after} (Color online) I-V curves obtained with
a memristor emulator wired as shown in the inset. The model given
by Eq. (\ref{Mmodel2}) was used with $\alpha=0$,
$\beta=62$k$\Omega/$V$\cdot$s, $V_T=1.75$V, $R_{min}=1$k$\Omega$
and $R_{max}=10$k$\Omega$. We used $V(t)=V_0\sin(2 \pi f t)$ as
the applied voltage with $V_0= 2.3$V and $f$'s as indicated on the
plot. The curves are noisy because of the small value of $R_{\,
0}=100\Omega$ used to find the current and of limited resolution
of our data acquisition system. We found that for the cases shown
in this plot the initial value of $R_M$ (in the present case equal
to $1$k$\Omega$) does not affect the long-time limit of the I-V
curves.}
 \end{center}
\end{figure}

As of today, memristors are not yet available on the market.
Therefore, in order to study memristor-based programmable
circuits, we have built a memristor emulator~\cite{Pershin2009-3}.
The latter is a simple electronic scheme (see Fig. \ref{fig2}a)
which can simulate a wide range of memristive systems. In
particular, we have recently used it to simulate the behavior of
synapses in simple neural networks~\cite{Pershin2009-3}. The main
element of the memristor emulator is a digital potentiometer whose
resistance is continuously updated by a microcontroller and
determined by pre-programmed equations of current-controlled or
voltage-controlled memristive systems. A general form of equations
describing a voltage-controlled memristive system is given by
Eqs.~(\ref{Condeq1}) and~(\ref{Condeq2}). These equations involve
a voltage drop on the memristor $V_M$ measured by the
analog-to-digital converter (ADC). A block scheme of the memristor
emulator operation algorithm is shown in Fig. \ref{fig2}b. The
algorithm's steps are self-explanatory.

In our experiments, we use an activation-type model of memristor
\cite{Pershin2009-2,Pershin2009-3} inspired by recent experimental
results \cite{Yang2008-1}. Within our model, $R_M=x$ and Eq.
(\ref{Condeq2}) is written as (with the resistance acquiring the
limiting values $R_{min}$ and $R_{max}$)

\begin{eqnarray}
\dot x&=&\left(\beta V_M+0.5\left( \alpha-\beta\right)\left[
|V_M+V_T|-|V_M-V_T| \right]\right) \nonumber\\ & &\times
\theta\left( x-R_{min}\right) \theta\left( R_{max}-x\right) \;
\label{Mmodel2},
\end{eqnarray}
where $\alpha$ and $\beta$ are constants defining memristance rate
of change below and above the threshold voltage $V_T$; $V_M$ is
the voltage on memristor and $\theta(\cdot)$ is the step function.
To test that our emulator does indeed behave as a memristor, we
have used the circuit shown in the inset of Fig. \ref{fig2after},
in which an ac voltage is applied to the memristor emulator
connected in series with a resistor which was used to determine
the current. The obtained current-voltage (I-V) curves, presented
in Fig. \ref{fig2after}, demonstrate typical features of
memristive systems such as pinched hysteresis loops and
frequency-dependent hysteresis.

\begin{table}[!t]
\begin{IEEEeqnarraybox}[
\IEEEeqnarraystrutmode\IEEEeqnarraystrutsizeadd{2pt}{1pt}]{v/l/v/c/v/c/v}
\IEEEeqnarrayrulerow\\ &\mbox{Parameter}&&\mbox{Real
memristor}&&\mbox{Memristor emulator}&\\ \IEEEeqnarraydblrulerow
\\ &\mbox{Resistance range}&&\mbox{Determined by the structure}&&\mbox{50$\Omega <R<10$k$\Omega$}&\\ \IEEEeqnarrayrulerow
\\ &\mbox{Discretization of $R$}&&\mbox{$R$ changes continuously}&&\mbox{256 steps}&\\ \IEEEeqnarrayrulerow
\\ &\mbox{Frequency}&&\mbox{Any}&&\mbox{$\lesssim 50$Hz}&\\ \IEEEeqnarrayrulerow
\\ &\mbox{Response}&&\mbox{Determined by the structure}&&\mbox{Determined by
pre$-$}&\\ &\mbox{}&&\mbox{}&&\mbox{programmed function}&\\
\IEEEeqnarrayrulerow
\\ &\mbox{Applied $V$}&&\mbox{Less than the breakdown }&&\mbox{0,+5V or -2.5,+2.5V}&\\
&\mbox{}&&\mbox{voltage of the structure}&&\mbox{}&\\
 \IEEEeqnarrayrulerow
\\ &\mbox{Supply $V$}&&\mbox{Not needed}&&\mbox{0,+5V or -2.5,+2.5V}&\\ \IEEEeqnarrayrulerow
\\ &\mbox{Max. continuous $I$}&&\mbox{Determined by the structure}&&\mbox{$\pm 11$mA}&\\\IEEEeqnarrayrulerow
\end{IEEEeqnarraybox}
\centering \caption{Comparison table of a solid-state memristor and
present version of memristor emulator.} \label{tabl1}
\end{table}

Table \ref{tabl1} shows a summary of main characteristics of a real
memristor and of the present version of memristor emulator. In the
case of memristor emulator, its characteristics are mainly limited
by the electronic component that we use and can be significantly
varied using different types of electronic components. As it is
shown in the Table \ref{tabl1}, the resistance of the present
memristor emulator can be tuned between 50$\Omega$ (this low
threshold is determined by unavoidable wiper resistance) and
10k$\Omega$ in 256 steps, as determined by the digital potentiometer
used. The ADC sampling frequency of 1kHz limits the characteristic
frequency of signals applied to memristor to approximately 50Hz (20
points per period are reasonably enough to simulate a real memristor
response). The voltage and current ratings of memristor emulator are
determined by absolute maximum ratings of AD5206 digital
potentiometer chip. By using a different hardware, the
characteristics of memristor emulator can be improved. For example,
the resolution can be easily increased to 1024 steps using a
different digital potentiometer, and characteristic operational
frequency can be as high as several tens of MHz using, e.g., a
modern 2 Giga-sample ultra-high-speed ADC (such as, for example,
ADC10D1000 offered by National Semiconductor). However, for the
purpose of demonstration, high resolution and high frequencies are
not required. This allows us to use inexpensive electronic
components.

Moreover, in order to have a good correspondence between the response of a real
memristor and that of a memristor emulator, it is extremely
important to have an appropriate device model. This model, in
terms of equations, is pre-programmed into the microcontroller and the
behavior of the memristor emulator would follow closely the given model (within the limits listed in
Table \ref{tabl1}). The model we employ (given by Eq.
(\ref{Mmodel2})) is quite simple. However, it
contains all important physics of solid-state memristive devices whose behavior derives from
activation-type processes.

\section{Applications} \label{sec3}
There is certainly a number of analog circuits where a memristor
can operate under the conditions described in Section
\ref{sec1}. Here, we show few examples of those applications.

\subsection{Programmable threshold comparator}

\begin{figure}[tb]
 \begin{center}
 \includegraphics[angle=0,width=6.5cm]{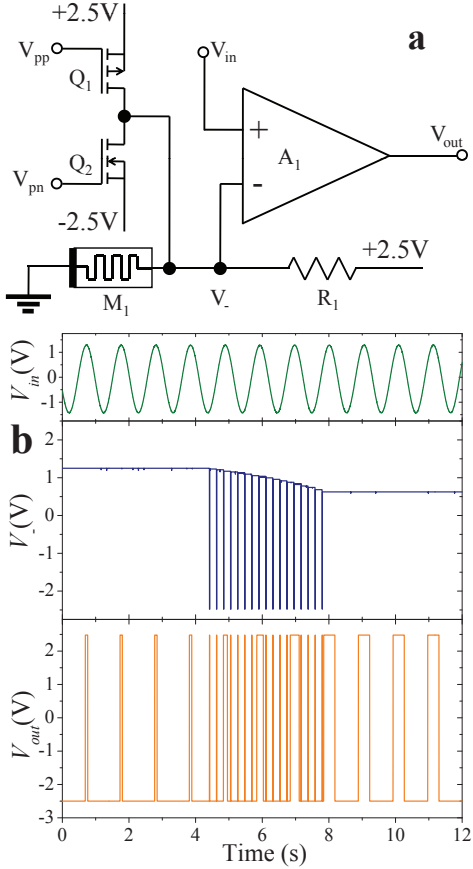}
\caption{\label{fig3} (Color online) {\bf a} Schematics of a
programmable threshold comparator. Here, M$_1$ is the memristor,
$R_1=10$k$\Omega$ and $A_1$ is an operational amplifier. In our
experiments, we used operational amplifier model TLV2770 (Texas
Instruments) that was powered by a dual polarity $\pm$2.5V power
supply. {\bf b} Programmable threshold comparator response to the
input voltage $V_{in}=V_0\sin (2\pi f t)$ with $V_0=1.3$V and
$f=1$Hz, and several negative programming pulses of 10ms width
applied in the time interval between 4 and 8 seconds. $V_{in}$ is
the input voltage applied to the positive input of the operational
amplifier, $V_-$ is the voltage on the negative input of the
operational amplifier, and $V_{out}$ is the signal at the output
of the operational amplifier. Each 10ms voltage pulse changes
$R_M$ by approximately 430$\Omega$ causing a gradual decrease of
the comparator threshold.}
 \end{center}
\end{figure}

Let us start by demonstrating memristor-based programmable analog
circuit operations with arguably the simplest case - a programmable
threshold comparator as shown in Fig. \ref{fig3}a. The design of
this circuit, as well as of all other circuits discussed below,
involves the memristor-based digital potentiometer block shown in
Fig. \ref{fig1}. In the analog mode of operation, both FETs are off.
In all our practical examples we build a scheme in such a way that
the maximum voltage drop on memristor is always smaller than the
threshold voltage of memristor which was selected to be equal to
1.75V. When possible, it is desirable to apply to memristor as low
voltages as possible in order to further reduce the slow change of
memristor state below its threshold.

In the programmable threshold comparator circuit, the comparator
threshold is determined by the voltage on the memristor given by
\begin{equation}
V_-=V_{cc}R_M/(R_M+R_1),
\end{equation}
where $V_{cc}=2.5$V is the power supply
voltage, and $R_1$ is the value of the resistance of the
``standard'' resistor. If the signal amplitude at the positive input $V_{in}$
exceeds $V_-$, then output signal $V_{out}$ is equal to the
saturation voltage of the operational amplifier (which in the present case is close to
$+2.5$V). In the opposite case, $V_{out}$ is close to $-2.5$V.

Fig. \ref{fig3}b shows the operation of the programmable threshold
comparator scheme when a sinusoidal voltage with an amplitude of
1.3V is applied to its input. At the initial moment of time, the
resistance of memristor is set to $R_M$=10k$\Omega$. That means
that $V_-=1.25$V and $V_{in}$ exceeds $V_-$ only for a short
period of time. As a result, we observe a set of narrow positive
pulses at the output $V_{out}$. The application of a train of
negative pulses in the time interval between 4 and 8 seconds
re-programs the memristor state by lowering the comparator
threshold. This can be observed in a smaller $V_{-}$ and wider
output pulses starting at $t=8$s.

\subsection{Programmable gain amplifier}

\begin{figure}[bt]
 \begin{center}
 \includegraphics[angle=0,width=7cm]{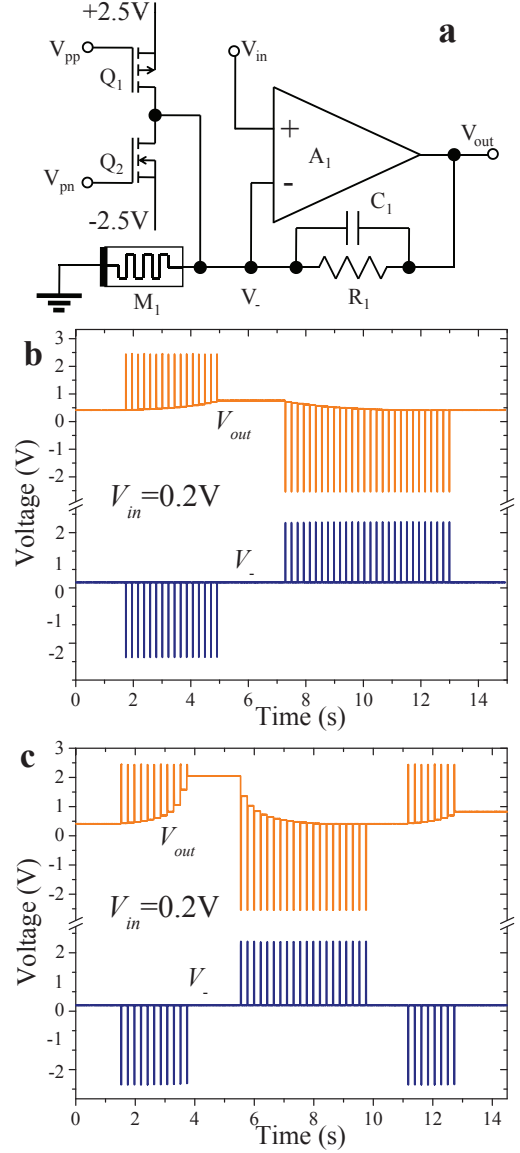}
\caption{\label{fig4} {\bf a}. Schematics of a programmable gain
amplifier with memristor. The operational amplifier A$_1$ is
connected in the standard non-inverting amplifier configuration with
a memristor M$_1$ replacing a resistor. Two FETs (Q$_1$ and Q$_2$)
are used to program the resistance of the memristor thus selecting
the amplifier gain. The capacitor C$_1=0.1\mu$F is used for noise
suppression. {\bf b}. Programming of gain using 10ms width pulses.
In these measurements, the input voltage $V_{in}=0.2$V is
permanently supplied while positive and negative pulses change the
state of memristor and correspondingly circuit's gain. As a result,
we observe a set of steps in the output signal. {\bf c}. Coarser
control of gain using 20ms width pulses.}
 \end{center}
\end{figure}

The next circuit we consider is a programmable gain
amplifier whose general scheme is shown in Fig. \ref{fig4}a. As in
the usual non-inverting amplifier configuration, the input signal
$V_{in}$ is applied to the positive input of the operational
amplifier A$_1$, while one of the two resistors, connected to the
negative input is substituted by a memristor M$_1$. The gain of
such amplifier ($R_1$ is the value of the resistance of the
``standard'' resistor)
\begin{equation}
V_{out}/V_{in}=1+R_1/R_M \label{gain}
\end{equation}
is determined by the value of memristance $R_M$ which can be
selected (programmed) between two limiting values $R_{min}$ and
$R_{max}$ using two field-effect transistors (FETs). The desirable
regime of operation that minimizes the voltage applied to the memristor is
$R_M \ll R_1$.

Figs. \ref{fig4}b and \ref{fig4}c demonstrate operation of the
programmable gain amplifier shown in Fig. \ref{fig4}a. The
amplifier's gain is controlled using pulses of constant width. The
pulse width is 10ms in Fig. \ref{fig4}b and 20ms in Fig.
\ref{fig4}c. In both cases, at the initial moment of time $t=0$,
the memristor is in its highest resistance state
$R_M=R_{max}=10k\Omega$ and, according to Eq. (\ref{gain}), the
circuit gain is about 2. Correspondingly, the input signal
$V_{in}=0.2$V results in $V_{out}=0.4$V at that time, as it can be
seen in Figs. \ref{fig4}b and \ref{fig4}c.

Application of pulses at $V_-$ changes the value of $R_M$ and,
correspondingly, the gain. Each negative pulse at $V_{-}$ decreases
the value of $R_M$ while each positive pulse increases $R_M$. Steps
in the output signal $V_{out}$ (separated by spikes due to the
voltage pulses during programming) correspond to different values of
circuit's gain (the input voltage $V_{in}$ is kept constant during
the experiment). For the selected memristor parameters, the circuit
gain changes approximately from  2 to 11. We demonstrate in Fig.
\ref{fig4}c that longer pulses produce larger changes in $R_M$
allowing for coarser control of the gain.

\subsection{Programmable switching thresholds Schmitt trigger}

\begin{figure}[tb]
 \begin{center}
 \includegraphics[angle=0,width=6.5cm]{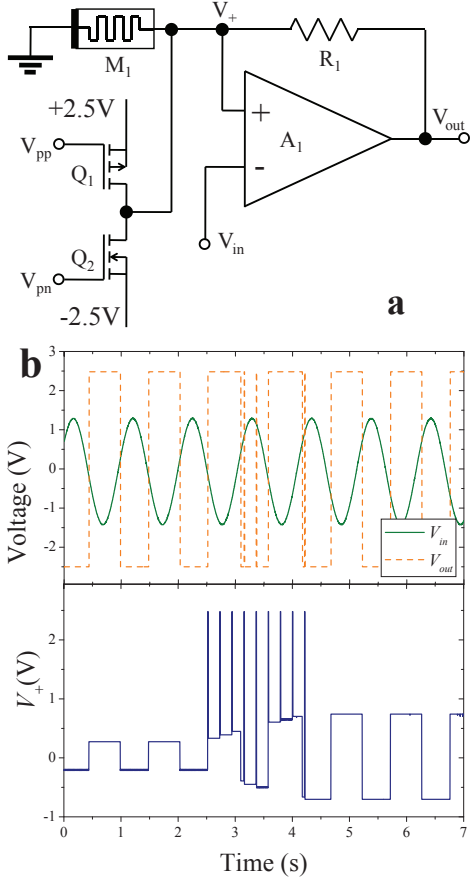}
\caption{\label{fig5} (Color online) {\bf a} Schematics of a
programmable switching thresholds Schmitt trigger with memristor.
Here, $R_1=10$k$\Omega$. {\bf b} Programmable switching thresholds
Schmitt trigger response to the input voltage $V_{in}=V_0\sin
(2\pi f t)$ with $V_0=1.3$V and $f=1$Hz, and several positive
programming pulses of 10ms width applied in the time interval
between 2 and 4 seconds.}
 \end{center}
\end{figure}

Fig. \ref{fig5}a shows schematics of a programmable switching
thresholds Schmitt trigger in the inverted configuration. This
circuit behaves as an inverted comparator with the switching
thresholds given by $\pm (R_M/R_1) V_{sat}$ where, for our circuit,
$V_{sat}=2.5$V. When we apply programming pulses to $M_1$, its
resistance $R_M$ changes as well as the switching thresholds of
Schmitt trigger. This type of behaviour is clearly seen in Fig.
\ref{fig5}b. Here, at the initial moment of time, the memristor is
in a low resistance state and, correspondingly, the switching
thresholds are low. Therefore, the switchings (changes of $V_{out}$
from -2.5V to 2.5V and vice-versa) occur close to the moment of time
when $V_{in}$ is close to zero as it follows from the positions of
the intersection points of $V_{in}$ with $V_{out}$ curves in the
upper panel of Fig. \ref{fig5}b.

A train of positive pulses applied to the memristor (see $V_+$ curve
in Fig. \ref{fig5}b) in a time interval between 2 and 4 seconds
increases $R_M$ and, consequently, the Schmitt trigger's switching
threshold. As a result, the upper panel in Fig. \ref{fig5}b
demonstrates that switching of $V_{out}$ occurs at different values
of $V_{in}$ when $t>4$s.

\subsection{Programmable frequency relaxation oscillator}

\begin{figure}[tb]
 \begin{center}
 \includegraphics[angle=0,width=7cm]{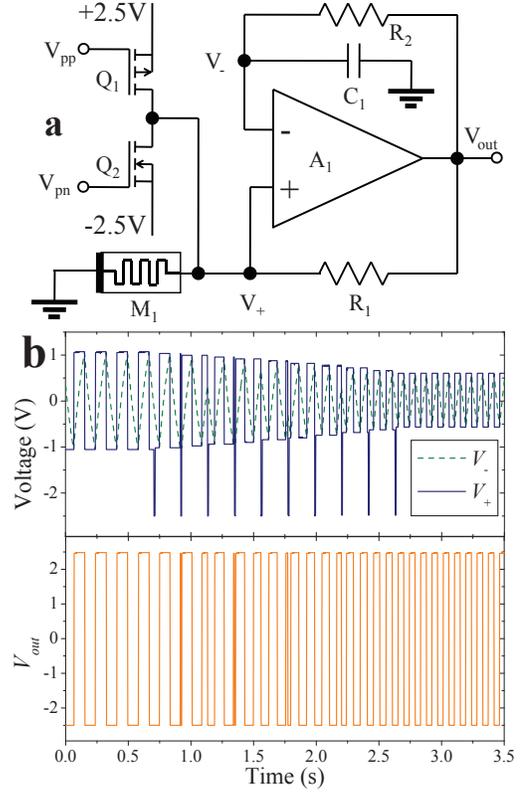}
\caption{\label{fig6} (Color online) {\bf a} Schematics of a
programmable frequency relaxation oscillator with memristor. Here,
$R_1=R_2=10$k$\Omega$ and $C=10\mu$F. {\bf b} Oscillating signals
in the different points of the programmable frequency relaxation
oscillator. The lower panel demonstrates an increase in the
oscillation frequency as $R_M$ is decreased by several negative
pulses applied to $V_+$.}
 \end{center}
\end{figure}

As a final example we consider a programmable frequency relaxation oscillator. This is schematically
shown in Fig.~\ref{fig6}a.
The relaxation oscillator is a well-known circuit which
automatically oscillates because of the negative feedback added to
a Schmitt trigger by an RC circuit. The period of oscillations is
determined by both the RC components and switching thresholds of the
Schmitt trigger. Therefore, in order to control the relaxation
oscillator frequency, we use a memristor-based
digital potentiometer to vary switching thresholds of the Schmitt
trigger, similarly to what is shown in Fig. \ref{fig5}a.

Fig. \ref{fig6}b demonstrates that a decrease of the memristor
resistance $R_M$ results in an increase of the relaxation
oscillator frequency. As it is demonstrated in the upper panel of
Fig. \ref{fig6}b, the decrease of the switching threshold results
in a faster capacitor charging time (because now the capacitor has
to charge to a smaller voltage). Via this mechanism, the memristor-based
digital potentiometer then determines the frequency of oscillations.

\section{Discussion and Conclusion} \label{sec4}

Having demonstrated the operation of memristor-based programmable
analog circuits, we would like to discuss certain practical
aspects of using the suggested approach with solid-state memristors that
are presently investigated. First of all, it should be mentioned that
although the rate of memristance change at low voltages is very
small as observed in certain experiments \cite{Yang2008-1}, it can
eventually lead to a drift  (increase or decrease depending on
particular application scheme) of $R_M$ at long times. Currently,
it is difficult to estimate this effect, in part because of the
lack of experimental data. However, this parasitic effect becomes
less important at low voltages applied to the memristor and, in
practice, can be corrected by periodic re-setting of $R_M$ or/and
circuit calibration. We also can not exclude the possibility that
the drift would become important only after several months or
years of device operation.

Another important point is the precision of the memristor state
programming. To program a desired value of memristance with a
given precision we should either have enough information on the
memristor operation model, parameters (and have
reproducibly-operating memristors), or provide pulses of precise
amplitude and duration. Alternatively we need to implement,
electronically, a suitable calibration procedure allowing for
correction of non-precise programming. Technical solutions for the
tasks mentioned above can certainly be found. Concerning
reproducibility of memristive behavior, experiments with TiO$_2$
thin films demonstrate a significant amount of noise in hysteresis
curves \cite{Yang2008-1}. Possibly, the resistance change effect
in colossal magnetoresistive thin films \cite{Liu2000-1} is more
suitable for analog-mode memristor applications.

In conclusion, we have suggested an approach for a practical
application of memristors in programmable analog circuits. To
demonstrate experimentally our approach, we have built several
memristor-based programmable analog circuits using a memristor
emulator. The latter is a scheme that uses inexpensive off-the-shelf components and can, therefore,
be built quite easily in any electronic lab. By applying a train of pulses, we have
achieved programming of different properties such as threshold,
gain and frequency. We have thus shown that a memristor with a control scheme provides a
simple realization of digital potentiometers and, therefore, can
find useful and broad-range applications in electronics.

\section*{Acknowledgment}

The authors are indebted to B. Mouttet for pointing out
Ref.~\cite{Liu2000-1} to us. This work has been partially funded
by the NSF grant No. DMR-0802830.

\ifCLASSOPTIONcaptionsoff
  \newpage
\fi



%

\bibliographystyle{IEEEtran}
\bibliography{IEEEabrv,memristor}
\end{document}